\documentclass[prl,twocolumn,showpacs]{revtex4-1}

\usepackage{amsfonts,amsmath,amssymb,graphicx}
\usepackage{amsfonts}
\usepackage{amsmath}
\usepackage{amssymb}
\usepackage{graphicx}

\def\kbar{\protect\@kbar}
\def\@kbar{\relax \bgroup
\def\@tempa{\hbox{\raise.73\ht0
\hbox to0pt{\kern.25\wd0\vrule width.5\wd0 height.1pt
depth.1pt\hss}\box0}}\mathchoice{\setbox0\hbox{$\displaystyle
k$}\@tempa}{\setbox0\hbox{$\textstyle
k$}\@tempa}{\setbox0\hbox{$\scriptstyle
k$}\@tempa}{\setbox0\hbox{$\scriptscriptstyle k$}\@tempa}\egroup}

\begin{document}

\title{\textbf{Sub-Fourier characteristics of a $\delta$-kicked rotor resonance} }
\author{  I. Talukdar, R. Shrestha, and G. S. Summy
} \affiliation{Department of Physics, Oklahoma State University,
Stillwater, Oklahoma 74078-3072, USA}

\begin{abstract}
\noindent We experimentally investigate the sub-Fourier behavior
of a $\delta$-kicked rotor resonance by performing a measurement
of the fidelity or overlap of a Bose-Einstein condensate (BEC)
exposed to a periodically pulsed standing wave. The temporal width
of the fidelity resonance peak centered at the Talbot time and
zero initial momentum exhibits an inverse cube pulse number
($1/N^{3}$) dependent scaling compared to a $1/N^{2}$ dependence
for the mean energy width at the same resonance. A theoretical
analysis shows that for an accelerating potential the width of the
resonance in acceleration space depends on $1/N^{3}$, a property
which we also verify experimentally. Such a sub-Fourier effect
could be useful for high precision gravity measurements.

\end{abstract}


\maketitle

The quantum $\delta$-kicked rotor (QDKR) has proved to be an
excellent testing ground for theoretical and experimental studies
of chaos in the classical and quantum domains \cite{qc}. An
experimental version of this system in the form of the kicked
particle is achieved by exposing cold atoms to $N$ pulses of an
off-resonant standing wave of light \cite{ao}. Ever since its
realization, the QDKR has continued to reveal a rich variety of
effects including dynamical localization \cite{dl}, quantum
accelerator modes \cite{qam,qam2}, quantum ratchets
\cite{Ratc,Ratch2} and quantum resonances \cite{ao,qr,horBP}. Such
resonances appear for pulses separated by rational fractions of a
characteristic time called the Talbot time and can be observed as
sharp peaks in the mean energy of the system \cite{Wimprl}. The
width of these peaks has been found to scale as $1/N^{2}$, a
sub-Fourier effect attributed to the non-linear nature of the QDKR
and explained using a semi-classical picture \cite{WimNL}. Away
from the resonances, dynamical localization sets in, characterized
by the quantum suppression of classical momentum diffusion beyond
a "quantum break time" \cite{dl}. This property, unique to quantum
dynamics in the chaotic regime, was utilized to discriminate
between two driving frequencies of the QDKR with sub-Fourier
resolution \cite{sfDL}.

\begin{figure}
\includegraphics[width=8cm]{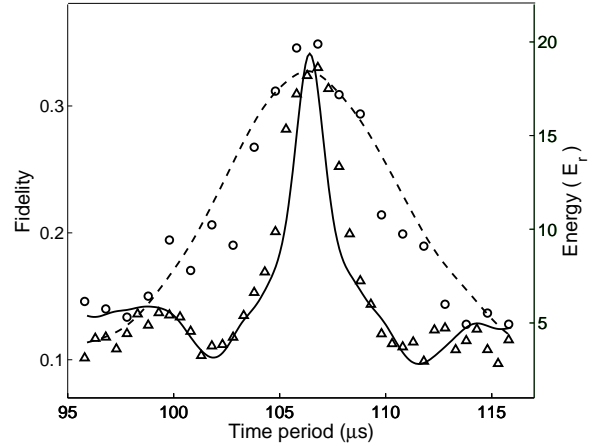}
\caption{Experimentally measured fidelity distribution (triangles)
due to 5 kicks of strength $\phi_{d}=0.8$ followed by a
$\pi$-phase shifted kick of strength $5\phi_{d}$. The mean energy
(circles) of the same 5 kicked rotor is shown for comparison.
Numerical simulations of the experiment for a condensate with
momentum width $0.06 \hbar G$ are also plotted for fidelity (solid
line) and mean energy (dashed line). The amplitude and offset of
the simulated fidelity were adjusted to account for the
experimentally imperfect reversal phase.}\label{fig1}
\end{figure}

High-precision measurements using quantum mechanical principles
have been carried with atom interferometers for many years
\cite{aint}. Such devices were used to determine the Earth's
gravitational acceleration \cite{grav}, fine structure constant
$\alpha$ \cite{fine}, and the Newtonian constant of gravity
\cite{newtG}. The promise of the QDKR as a candidate for making
these challenging measurements has begun to be realized
\cite{Ton}. Recently a scheme was proposed for measuring the
overlap or fidelity between a near-resonant $\delta$-kicked rotor
state and a resonant state via application of a tailored pulse at
the end of a rotor pulse sequence \cite{Fid}. It predicted a
$1/N^{3}$ scaling of the temporal width of the fidelity peak. In
this paper we report on the observation of such fidelity resonance
peaks and their sub-Fourier nature. Figure \ref{fig1} illustrates
a plot of the fidelity (fraction of atoms in the zeroth order
momentum state) vs pulse period obtained by the application of an
overlap pulse at the end of five rotor kicks. For comparison we
also plot the mean energy of the rotor sequence. It can be seen
that even for relatively few kicks the fidelity peak is
significantly narrower. We also investigated the sensitivity of
this fidelity resonance to an accelerating rotor. As will be seen
our calculations indicate that the width of the fidelity peak vs
acceleration decreases at a sub-Fourier rate of $1/N^{3}$. We
confirm this result with experiments.

The dynamics of a periodically kicked atom in the presence of a
linear potential is described by the quantum $\delta$-kicked
accelerator (QDKA) Hamiltonian
\begin{equation}
\hat{H}=\frac{\hat{P}^{2}}{2} +\frac{\eta}{\tau}\hat{X}
+\phi_{d}\cos(\hat{X})\sum_{t=1}^{N}\delta (t'-t\tau).  \label{Ham}
\end{equation}
$\hat{P}$ is the momentum (in units of two photon recoils, $\hbar
G$) that an atom of mass $M$ acquires from short, periodic pulses
of a standing wave with a grating vector $G=2\pi/\lambda_{G}$
($\lambda_{G}$ is the wavelength of the standing wave). $\hat{X}$
is position in units of $G^{-1}$ and $\eta=Mg'T/\hbar G$, $g'$
being its acceleration between pulses separated by $T$, the pulse
period. $\phi_{d}=\Omega^{2}\Delta t/8\delta_{L}$ represents the
kicking strength of a pulse of length $\Delta t$, $\Omega$ is the
Rabi frequency and $\delta_{L}$ the detuning of the kicking laser
from the atomic transition. $t'$ is the continuous time variable
and $\tau=2\pi T/T_{1/2}$ is the scaled pulse period. For the case
$\eta=0$, Eq.\thinspace(\ref{Ham}) reduces to the familiar QDKR
Hamiltonian. Primary quantum resonances are seen for pulses
separated by integer multiples of the half-Talbot time,
$T_{1/2}=2\pi M/\hbar G^{2}$ or $\tau=2\pi$. Adjacent momentum
orders evolve integer multiples of $2\pi$ in phase during this
time period resulting in a quadratic growth in the rotor mean
energy, $\langle E\rangle=2E_{r}\phi_{d}^{2}N^{2}$, where
$E_{r}=\hbar^{2}G^{2}/8M$ is the photon recoil energy. The width
of the mean energy around the resonance time was found to decrease
as $1/(N^{2}\phi_{d})$ \cite{WimNL,horBP}.

In order to demonstrate the role of the relative phase deviations
of the contributing momentum states near such a resonance, a
\textquotedblleft fidelity\textquotedblright test for the QDKR was
proposed in Ref.\thinspace \cite{Fid}. In this scheme, a kick reversed in
phase by $\pi$ and carrying a strength of $N\phi_{d}$ is applied
at the end of the $N$ rotor kicks. The fidelity is then defined as
$F= |\langle \beta|U_{r}U^{N}|\beta \rangle |^{2}$ where
$U=\exp(-i\frac{\tau}{2}\hat{P}^{2})\exp[-i\phi_{d}\cos(\hat{X})]$
describes the one period evolution,
$U_{r}=\exp[iN\phi_{d}\cos(\hat{X})]$ is the overlap pulse and
$\beta$ is the fractional part of the momentum. $F$ therefore
gives the probability of the revival of the initial state and is
measured by the fraction of atoms which have returned to the
initial zero momentum state. Near a resonance at the Talbot time,
$\tau=4\pi$, the fidelity is \cite{Fid},
\begin{equation}
F(\epsilon,\beta=0)\simeq
J^{2}_{0}\left(\frac{1}{12}N^{3}\phi_{d}^{2}\epsilon
\right)\label{FwN}
\end{equation}
where $\epsilon=\tau-4\pi$. The width of such a peak in $\epsilon$
therefore changes as $1/(N^{3}\phi_{d}^{2})$, displaying a
stronger sub-Fourier dependence on the number of kicks than the
mean energy.

Our experiment is performed by producing a BEC of 20000 Rb87 atoms
in the $5S_{1/2}, F=1, m_{F}=0$ level in an optical trap
\cite{qam2,chap}. After being released from the trap, the
condensate is exposed to a horizontal standing wave created by two
beams of wavelength $\lambda=$ 780 nm light detuned 6.8 GHz to the
red of the atomic transition. The wave vector of each beam was
aligned $\theta=52^{\text{\textrm{o }}}$ to the vertical. This
created a horizontal standing wave with a wavelength of
$\lambda_{G}=\lambda/2\sin \theta$ and a corresponding Talbot time
of 106.5 $\mu s$. Two acousto-optic modulators (AOMs) controlled
the pulse lengths as well as the relative frequencies of the
kicking beams enabling the control of the acceleration and initial
momentum of the standing wave with respect to the condensate. The
kicking pulse length was 0.8 $\mu s$ with a $\phi_{d}\approx$ 0.6.
For the last kick the phase of one of the AOMs RF driving signal
was changed by $\pi$ which shifted the standing wave by half a
wavelength. In order to keep this final overlap pulse within the
Raman-Nath regime we varied the intensity rather than the pulse
length to create a kick strength of $N\phi_{d}$. This was done by
adjusting the amplitudes of the RF waveforms driving the kicking
pulse. Dephasing due to experimental instabilities made the
reversal process inconsistent for $N>$6. To reduce this effect, a
$\pi$-shifted kicking sequence was adopted where each kick in the
QDKR was shifted by half a wavelength with respect to the
previous. This decreased the Talbot time by half and led to much
improved results for larger number of kicks. Following the entire
kicking sequence we waited 8 ms for the different momentum orders
to separate before the atoms were absorption imaged.

\begin{figure}
\includegraphics[width=8cm]{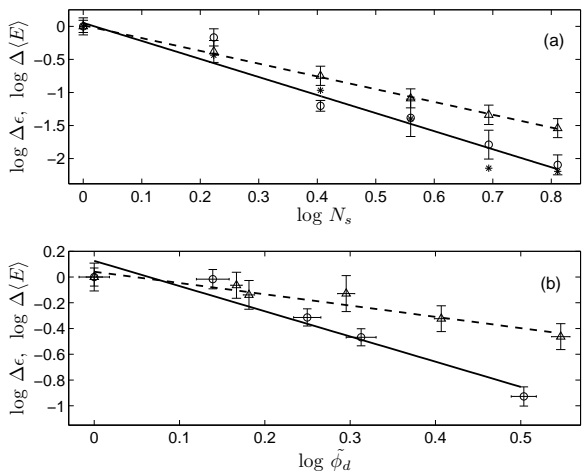}
\caption{Experimentally measured fidelity(circles) and mean
energy(triangles) widths(FWHM) as a function of (a) the number of
pulses, and (b) the kicking strength $\tilde{\phi_{d}}$ scaled to
$\phi_{d}$ of the first data point. In (a), the data are for 4 to
9 kicks in units normalized to the 4$^{\textrm{th}}$ kick. Error
bars in (a) are over three sets of experiments and in (b)
1$\sigma$ of a Gaussian fit to the distributions. Dashed lines are
linear fits to the data. Stars are numerical simulations for an
initial state with a momentum width of $0.06 \hbar G$.
}\label{fig2}
\end{figure}

From the time of flight images fidelity $F$ is measured as the
fraction of atoms which have reverted back to the zeroth order
momentum state, that is $F=P_{0}/\sum_{n} P_{n}$ where $P_{n}$ is
the number of atoms in the $n^{\textrm{th}}$ momentum order. To
facilitate the analysis of the data, all of the resonance widths
($\delta\epsilon$) were scaled to that at a reference kick number
of $N=4$. That is we define a scaled fidelity width
$\Delta{\epsilon}=\delta\epsilon/\delta\epsilon_{N=4}$ for each
scaled kick number $N_{s}=N/4$ and recover $
\log \Delta{\epsilon}=-3\log N_{s}$ using
Eq.\thinspace(\ref{FwN}). For each kick, a scan is performed
around the resonance time. To ensure the best possible fit of the
central peak of the fidelity spectrum to a gaussian, the time is
scanned between values which make the argument of $J_{0}^{2}$ of
Eq.\thinspace(\ref{FwN}) $\approx2.4$ so that the first side lobes
are only just beginning to appear. Figure \ref{fig2}(a) plots the
logarithm of the full width at half maximum for 4 to 9 kicks
scaled to the fourth kick. A linear fit to the data gives a slope
of $-2.73\pm0.19$ giving a reasonable agreement with the predicted
value of -3 within the experimental error. As seen in the same
figure, the results are close to the numerical simulations which
take into account the finite width of the initial state of
$0.06\hbar G$ \cite{Ratch2}.
We also compared the resonance widths of the kicked rotor mean energy $\langle E\rangle$
to that of the fidelity widths. As in the fidelity, the plotted
values $\Delta\langle E\rangle$ have been normalized to that of
the fourth kick. On the log scale, the width of each peak gets
narrower with the kick number with a slope of $-1.93\pm0.21$ (Fig.\thinspace
\ref{fig2}(a)) in agreement with previous results
\cite{WimPRA,horBP}. As a further test of Eq. (\ref{FwN}), the
variation in the widths of the fidelity and mean energy peaks were
studied as a function of $\phi_{d}$. Figure \ref{fig2}(b) shows
the fidelity width changing with a slope of $-1.96\pm0.3$, close
to the predicted value of -2. This is again a faster scaling
compared to the mean energy width which decreases with a slope of
$-0.88\pm0.24$ (the theoretical value being -1).

\begin{figure}[hbp]
\includegraphics[width=7cm]{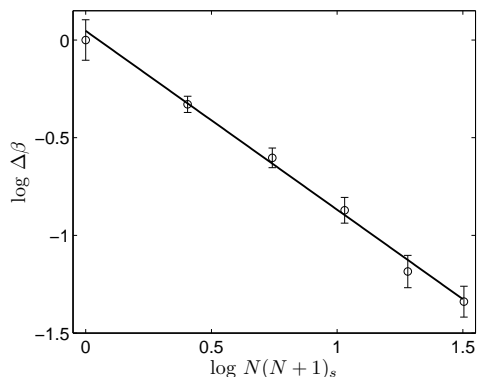}
\caption{Variation of the fidelity peak width around $\beta$=0 as
a function of kick number $N(N+1)_{s}=N(N+1)/20$ scaled to the
4$^{\textrm{th}}$ kick. The straight line is a linear fit to the
data with a slope of $-0.92\pm0.06$. Errorbars as in Fig.
2(b).}\label{fig4}
\end{figure}

The resonances studied here appear for pulses separated by the
Talbot time and an initial momentum state of $\beta=0$. Away from
this resonant $\beta$, phase changes in the amplitudes of the
different momentum orders lead to a fidelity which depends on the
initial momentum as, $F(\epsilon=0,\beta)=J^{2}_{0}(2\pi
\phi_{d}N(N+1)\beta)$ \cite{Fid}. The peak width in $\beta$ space
is thus expected to change as $1/[N(N+1)]$ around
$\beta=0$, as against a $1/N$ scaling of the mean energy width
\cite{horBP}. To verify this, the initial momentum of the
condensate with respect to the standing wave was varied
and the kicking sequence applied. The experimentally measured
widths $\Delta\beta=\delta\beta/\delta\beta_{N=4}$ in Fig.
\ref{fig4} display a scaling of
$\Delta\beta\varpropto[N(N+1)]^{-0.92}$ close to the theoretical
value.

\begin{figure}
\includegraphics[width=8cm]{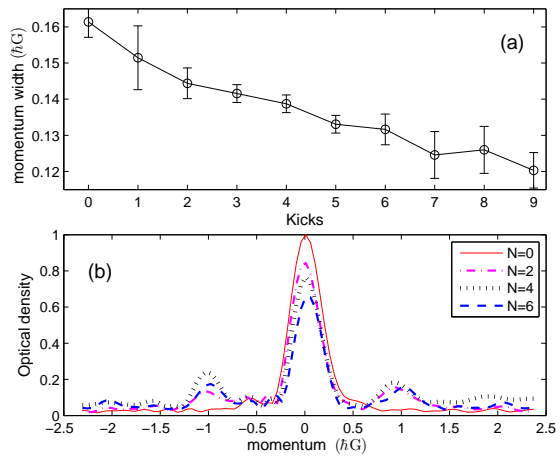}
\caption{(a)Momentum width of the reversed zeroth order state as a
function of kick number. Errorbars are an average over three
experiments. (b)Optical density plots for the initial state
(red,solid) and kick numbers 2 (magenta,dot-dashed),4
(black,dotted), and 6 (blue,dashed) after summation of the
time-of-flight image along the axis perpendicular to the standing
wave.}\label{fig5}
\end{figure}

For an initial state $|\beta+n\rangle$, the wave function acquires
a non-zero phase during the free evolution even at the Talbot
time. Therefore the final kick performs a velocity selective
reversal, preferentially bringing back atoms closer to an initial
momentum of $\beta=0$. This is similar to the time-reversed
Loschmidt cooling process proposed in Refs. \cite{Losch1,Losch2},
although in that technique a forward and reverse path situated on
either side of the resonant time was used in order to benefit from
the chaotic dynamics. To observe this effect the current scheme
offers experimental advantage in terms of stability due to the
reduced length of the pulse sequence. Here, only a single pulse
performs the velocity selection at the end, whereas in the
Loschmidt technique $N$ phase reversed kicks separated by a finite
time are used. This \textquotedblleft cooling\textquotedblright
effect is demonstrated in Fig. \ref{fig5} which shows the
reduction in the momentum distribution width of the reversed
zeroth order state from the initial condensate for a fidelity
measurement on resonance. Figure \ref{fig5}(b) depicts the
profiles of this state for $N$=0, 2, 4 and 6. Accompanying the
momentum width decrease is a drop in the peak height. Our own
simulations and the results of Ref. \cite{Losch2} predict that for
the case of a non-interacting condensate this should remain
constant. In addition to interactions we expect experimental
imperfections in the fidelity sequence to play a role in the
smaller peak densities with increasing kick numbers. We performed
the same experiment 4.5 ms after the BEC was released from the
trap when the mean field energy had mostly been transformed to
kinetic energy in the expanding condensate. A similar reduction in
the momentum width of the reversed state along-with a decrease in
the peak density was observed.

\begin{figure}[tbp]
\includegraphics[width=7cm]{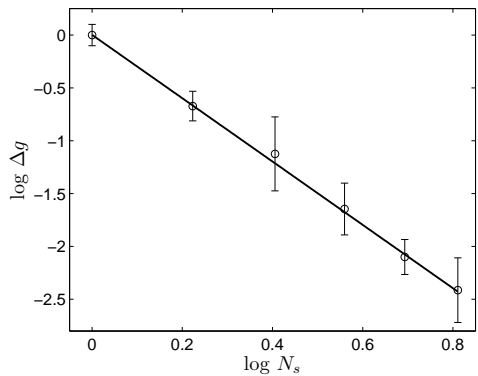}
\caption{Dependence of the acceleration resonance peak width as a
function of the kick number in units scaled to the
4$^{\textrm{th}}$ kick. Error bars are over three sets of
experiments.}\label{fig3}
\end{figure}

We now investigate the behavior of fidelity in the presence of
acceleration. The state of the QDKA Eq. (\ref{Ham}) after $N$
kicks is $|\psi(N\tau)\rangle=\sum_{n}c_{n}|n+\beta\rangle$ where
$n$ is the integer part of momentum $\hat{P}$. The expansion
coefficients $c_{n}$ are $c_{n}(\epsilon,\beta,\eta)=\langle
n+\beta|\hat{U_{g_{N}}}...\hat{U_{g_{2}}}\hat{U_{g_{1}}}|\beta\rangle$.
$\hat{U_{g_{t}}}=\exp[-i\frac{\tau}{2}(\hat{p}+t\eta+\frac{\eta}{2})^{2}]\exp[-i\phi_{d}\cos(\hat{X})]$
is the $t^{\textrm{th}}$ kick evolution operator in the freely
falling frame obtained after a gauge transformation of the
Hamiltonian (\ref{Ham}) which restores the conservation of
\textit{quasimomentum} $\beta$ \cite{fgr}. Close to the
resonances, we have
$F(\epsilon,\beta,\eta)\simeq|\sum_{n}J^{2}_{n}(N\phi_{d})\exp(-i\Theta_{n})|^{2}$,
where $ \Theta_{n}=\frac{\partial\theta_{n}}{\partial\epsilon}|\epsilon+
\frac{\partial\theta_{n}}{\partial\beta}|\beta+\frac{\partial\theta_{n}}{\partial\eta}|\eta
$ describes the effect of deviations from resonance on the
coefficients $c_{n}$. Using a procedure detailed in Ref.
\cite{Fid}, one can show that
$\frac{\partial\theta_{n}}{\partial\epsilon}|_{(\epsilon=\beta=\eta=0)}
=\frac{\frac{\partial c_{n}}{\partial\eta}|}{ic_{n}(0,0,0)}=-4\pi
nN^{2}/3,$ where we have kept terms in $N^{2}$. Finally we arrive
at the fidelity in the presence of acceleration,
\begin{equation}
F(\eta,\epsilon=\beta=0)=J^{2}_{0}\left(\frac{4\pi}{3}N^{3}\phi_{d}\eta\right).\label{fid-g}
\end{equation}
Thus the width of such a peak centered at the resonant zero
acceleration should drop as $1/N^{3}$. In order to verify the above result, the standing wave was
accelerated during the application of the pulses. This
acceleration was scanned across the resonant zero value and
readings of the fidelity collected. Figure \ref{fig3} plots the
experimental data for 4 to 9 kicks, where the widths of the peaks
decrease with a slope of $-3.00\pm0.23$ in excellent agreement
with the theory.

In conclusion, we performed experimental measurements of the
fidelity widths of a $\delta$-kicked rotor state near a quantum
resonance. The width of these peaks centered at the Talbot time
decreased at a rate of $N^{-2.73}$ comparable to the predicted
exponent of $-3$. By comparison, the mean energy widths was found
to reduce only as $N^{-1.93}$. Furthermore, the fidelity peaks in
momentum space changed as $(N(N+1))^{-0.92}$, also consistent with
theory. The reversal process used in the fidelity experiments was
found to lead to a cooling effect, whereby the momentum
distribution of the final zeroth order state decreased by 25$\%$
from the initial width at the end of a 9$^{\textrm{th}}$ reversal
kick, as a result of the velocity selection by the final pulse.
The sub-Fourier dependencies of the mean energy and fidelity
observed here are characteristic of the dynamical quantum system
that is the QDKR \cite{Fid}. The narrower resonances of the
fidelity scheme could be exploited in locating the resonance
frequency with a resolution below the limit imposed by the fourier
relation. This could be used to determine the photon recoil
frequency ($\omega_{r}=E_{r}/\hbar$) which together with the
photon wavelength enables measurement of the fine structure
constant with a high degree of precision \cite{Ton,fine}. We also
demonstrated a $N^{-3}$ dependence of the resonance width in
acceleration space in accordance with the extended theory. The
sensitivity of an atom interferometer based gravimeter scales as
the square of the loop time, hence the pursuit of large area
interferometers to improve accuracy \cite{grav}. The possibilities
offered by a process like the fidelity measure which is responsive
to gravity with the cube of the 'time' $N$ becomes evident here.
With future refinements this scheme could therefore serve as a
highly sensitive measure of the local gravity.

This work was supported by the NSF under Grant No. PHY-0653494.

\end{document}